\documentclass[conference,10pt,letterpaper]{IEEEtran}
\newif\ifextended
\extendedtrue

\newif\ifdraft
\draftfalse

\usepackage[bookmarks=true,breaklinks=true,colorlinks,linkcolor=black,citecolor=blue,urlcolor=black]{hyperref}
\usepackage[compress]{cite}

\usepackage{listings}
\usepackage{fancyhdr}

\usepackage{datetime}

\usepackage{amsmath,amssymb,amsfonts}
\usepackage{algorithmic}
\usepackage{graphicx}
\usepackage{textcomp}
\usepackage{xcolor}
\usepackage{tikz}
\usetikzlibrary{calc}
\usetikzlibrary{fit}

\usepackage[utf8]{inputenc}
\usepackage[T1]{fontenc}

\usepackage{booktabs} 
\usepackage{setspace}
\usepackage[italic]{mathastext}
\usepackage{array}
\usepackage{titlesec}
\usepackage[normalem]{ulem}
\usepackage{multirow}
\usepackage{multicol}
\usepackage{color}
\usepackage[font={scriptsize, stretch=0.8}]{caption}
\usepackage{float}
\usepackage[font={small}]{subcaption}
\usepackage[linesnumbered,ruled]{algorithm2e}
\usepackage{makecell}

\usepackage{pifont}
\usepackage[]{microtype}
            
\usepackage{marginnote} 
\usepackage{pifont}
\usepackage{titlesec}

\usepackage[acronym,nonumberlist,nowarn]{glossaries}
    \glsdisablehyper
     \newacronym{ADI}{ADI}{Alternating Direction Implicit}
\newacronym{AGU}{AGU}{Address Generation Unit}
\newacronym[longplural={Access History Tables}]{AHT}{AHT}{Access History Table}
\newacronym{AHT-R}{AHT-R}{Access History Table - Read}
\newacronym{AHT-W}{AHT-W}{Access History Table - Write-Back}
\newacronym{AIP}{AIP}{Access Interval Predictor}
\newacronym{AL}{AL}{Added Latency for column accesses}
\newacronym{ASIC}{ASIC}{Application Specific Integrated Circuit}
\newacronym{AVX}{AVX}{Advanced Vector Extensions}
\newacronym[longplural={Basic Block Vectors}]{BBV}{BBV}{Basic Block Vector}
\newacronym{BHT}{BHT}{Branch History Table}
\newacronym{HBM}{HBM}{Hybrid Bandwidth Memory}
\newacronym{DDR4}{DDR4}{Double-Data Rate 4}
\newacronym{MIMD}{MIMD}{multiple-instruction multiple-data}
\newacronym{SIMT}{SIMT}{single-instruction multiple-thread}
\newacronym{SLP}{SALP}{subarray-level parallelism}

\newacronym{NN}{NN}{Neural Network}
\newacronym{BTB}{BTB}{Branch Target Buffer}
\newacronym{CAS}{CAS}{Column Address Strobe}
\newacronym{AMAT}{AMAT}{Average Memory Access Time}
\newacronym{CCD}{CCD}{Column to Column Delay}
\newacronym{AI}{AI}{Arithmetic Intensity}
\newacronym[longplural={Columnar Database Systems}]{CDBS}{CDBS}{Columnar Database System}
\newacronym[longplural={Chip Multiprocessors}]{CMP}{CMP}{Chip Multiprocessor}
\newacronym{CMT}{CMT}{Chip Multithreading}
\newacronym[longplural={Coarse-Grain Reconfigurable Arrays}]{CGRA}{CGRA}{Coarse-Grain Reconfigurable Array}
\newacronym{C-RAM}{C-RAM}{Computational-RAM}
\newacronym{CWD}{CWD}{Column Write Delay}
\newacronym{DBI}{DBI}{Dynamic Binary Instrumentation}
\newacronym[longplural={Database Management Systems}]{DBMS}{DBMS}{Database Management System}
\newacronym{DDR}{DDR}{Double Data Rate}
\newacronym{DEWP}{DEWP}{Dead Line and Early Write-Back Predictor}
\newacronym{DRAM}{DRAM}{Dynamic Random Access Memory}
\newacronym{DSBP}{DSBP}{Dead Sub-Block Predictor}
\newacronym{DSM}{DSM}{Decomposed Storage Manage}
\newacronym{FAW}{FAW}{Four row Activation Window}
\newacronym{FP}{FP}{Floating-Point}
\newacronym{EDP}{EDP}{Energy-Delay Product}
\newacronym{FCFS}{FCFS}{First-Come First-Serve}
\newacronym{FIFO}{FIFO}{Fist In, First Out}
\newacronym{FSM}{FSM}{Finite State Machine}
\newacronym{FPGA}{FPGA}{Field-Programmable Gate Array}
\newacronym[longplural={Functional Units}]{FU}{FU}{Functional Unit}
\newacronym{GAg}{GAg}{Global Adaptive branch prediction using one Global PHT}
\newacronym{GAs}{GAs}{Global Adaptive branch prediction using per-Set PHT}
\newacronym{GCC}{GCC}{GNU Compiler Collection}
\newacronym{GEMS}{GEMS}{General Execution-driven Multiprocessor Simulator}
\newacronym[longplural={General Purpose Processors}]{GPP}{GPP}{General Purpose Processor}
\newacronym{HMC}{HMC}{Hybrid Memory Cube}
\newacronym{HIVE}{HIVE}{HMC Instruction Vector Extensions}
\newacronym{IATAC}{IATAC}{Inter-Access Time per Access Count}
\newacronym{ILP}{ILP}{Instruction Level Parallelism}
\newacronym{IPC}{IPC}{Instructions per Cycle}
\newacronym{ISA}{ISA}{Instruction Set Architecture}
\newacronym{IRAM}{IRAM}{Intelligent RAM}
\newacronym{KIPS}{KIPS}{Kilo Instructions per Second}
\newacronym{LDS}{LDS}{Linked Data Structure}
\newacronym{LFSR}{LFSR}{Linear Feedback Shift Register}
\newacronym{LFMR}{LFMR}{Last-to-First Miss-Ratio}
\newacronym{MLP}{MLP}{Memory-Level Parallelism}

\newacronym{LLC}{LLC}{last-level cache}
\newacronym{LRU}{LRU}{Least Recently Used}
\newacronym{LSU}{LSU}{Load Store Unit}
\newacronym{LTP}{LTP}{Last-Touch Predictor}
\newacronym{LvP}{LvP}{Live-time Predictor}
\newacronym{LWP}{LWP}{Last Write Predictor}
\newacronym{MARSS}{MARSS}{Micro Architectural and System Simulator}
\newacronym{McPAT}{McPAT}{Multi-core Power, Area, and Timing}
\newacronym{MIPS}{MIPS}{Microprocessor without Interlocked Pipeline Stages}
\newacronym{MOB}{MOB}{Memory Order Buffer}
\newacronym{MMU}{MMU}{memory management unit}

\newacronym{MPKI}{MPKI}{Misses per Kilo-Instruction}
\newacronym{MSHR}{MSHR}{Miss-Status Handling Registers}
\newacronym{NAS}{NAS}{Numerical Aerodynamic Simulation}
\newacronym{NDP}{NDP}{Near-Data Processing}
\newacronym{NMP}{NMP}{Near Memory Processor}
\newacronym{NoC}{NoC}{Network-on-Chip}
\newacronym{NPB}{NPB}{NAS Parallel Benchmark}
\newacronym{NUCA}{NUCA}{Non-Uniform Cache Architecture}
\newacronym{NUMA}{NUMA}{Non-Uniform Memory Access}
\newacronym{OoO}{OoO}{Out-of-Order}
\newacronym{OpenMP}{OpenMP}{Open Multi-Processing}
\newacronym{OS}{OS}{Operating System}
\newacronym{PAg}{PAg}{Per-address Adaptive branch prediction using one Global PHT}
\newacronym{PAs}{PAs}{Per-address Adaptive branch prediction using per-Set PHT}
\newacronym{PC}{PC}{Program Counter}
\newacronym[longplural={Pointer-Chasing Engines}]{PCE}{PCE}{Pointer-Chasing Engine}
\newacronym{PCM}{PCM}{Performance Counter Monitor}
\newacronym{PIM}{PiM}{Processing-in-Memory}
\newacronym[longplural={Pattern History Tables}]{PHT}{PHT}{Pattern History Table}
\newacronym{RAPL}{RAPL}{Running Average Power Limit}
\newacronym{RAS}{RAS}{Row Address Strobe}
\newacronym{RAT}{RAT}{Registers Alias Table}
\newacronym{RC}{RC}{Row Cycle}
\newacronym{RCD}{RCD}{RAS to CAS Delay}
\newacronym[longplural={Row-based Database Systems}]{RDBS}{RDBS}{Row-based Database System}
\newacronym{ROB}{ROB}{Reorder Buffer}
\newacronym{RRD}{RRD}{Row to Row activation Delay}
\newacronym{DNN}{DNN}{Deep Neural Network}
\newacronym{ANN}{ANN}{Artificial Neural Network}

\newacronym{pim}{PiM}{Processing-in-Memory}
\newacronym{pid}{PiD}{Processing-in-DRAM}
\newacronym{pnm}{PnM}{Processing-near-Memory}
\newacronym{pnd}{PnD}{Processing-near-DRAM}
\newacronym{pum}{PuM}{Processing-using-Memory}
\newacronym{pud}{PuD}{Processing-using-DRAM}

\newacronym{RP}{RP}{Row Precharge}
\newacronym{RTL}{RTL}{Register Transfer Level}
\newacronym{RTP}{RTP}{Read To Precharge}
\newacronym{RVU}{RVU}{Reconfigurable Vector Unit}
\newacronym{SDP}{SDP}{Skewed Dead-Block Predictor}
\newacronym{SESC}{SESC}{Superescalar Simulator}
\newacronym{SSE}{SSE}{Streaming SIMD Extensions}
\newacronym{SFP}{SFP}{Spatial Footprint Predictor}
\newacronym{SiNUCA}{SiNUCA}{Simulator of Non-Uniform Cache Architectures}
\newacronym{SMT}{SMT}{Simultaneous Multi-Threading}
\newacronym{SIMD}{SIMD}{single-instruction multiple-data}
\newacronym{SPEC}{SPEC}{Standard Performance Evaluation Corporation}
\newacronym{SoC}{SoC}{System-on-Chip}
\newacronym{SPP}{SPP}{Spatial Pattern Predictor}
\newacronym{SRAM}{SRAM}{Static Random Access Memory}
\newacronym{SSOR}{SSOR}{Symmetric Successive Over-Relaxation}
\newacronym{SSV}{SSV}{Search Set Vector}
\newacronym{TLB}{TLB}{Translation Look-aside Buffer}
\newacronym{TLP}{TLP}{Thread Level Parallelism}
\newacronym[longplural={Through-Silicon Vias}]{TSV}{TSV}{Through-Silicon Via}
\newacronym{VWQ}{VWQ}{Virtual Write Queue}
\newacronym{WTR}{WTR}{Write Recovery time}
\newacronym{WR}{WR}{Write To Read delay time}
\newacronym{TRA}{TRA}{triple row activation}

\definecolor{denim}{rgb}{0.08, 0.38, 0.74}
\definecolor{darkolivegreen}{rgb}{0.33, 0.42, 0.18}
\definecolor{dgreen}{rgb}{0.00, 0.75, 0.00}
\definecolor{darkpink}{rgb}{0.88, 0.28, 0.54}
\definecolor{forestgreen}{rgb}{0.0, 0.27, 0.13}
\definecolor{amber}{rgb}{1.0, 0.49, 0.0}
\definecolor{lightyellow}{rgb}{0.980, 0.956, 0.623}
\definecolor{lightblue}{rgb}{0.980, 0.956, 0.623}
\definecolor{darkamber}{rgb}{0.5, 0.19, 0.0}
\definecolor{dkgreen}{rgb}{0,0.6,0}
\definecolor{gray}{rgb}{0.5,0.5,0.5}
\definecolor{lightmauve}{rgb}{0.68,0.4,0.92}
\definecolor{chocolate}{rgb}{0.48, 0.25, 0.0}
\definecolor{dollarbill}{rgb}{0.52,0.73,0.4}
\definecolor{dkdkgreen}{rgb}{0,0.45,0}
\definecolor{gfored}{rgb}{0.580, 0.050, 0.211}
\definecolor{darkwarmgray}{rgb}{0.15, 0.050, 0.05}
\definecolor{ups-truck}{rgb}{0.53, 0.28, 0.21}

    \usepackage[colorinlistoftodos,prependcaption,textsize=tiny]{todonotes}
\ifdraft    
    
    \paperwidth=\dimexpr \paperwidth + 4cm\relax
    \oddsidemargin=\dimexpr\oddsidemargin + 2cm\relax
    \evensidemargin=\dimexpr\evensidemargin + 2cm\relax
    \marginparwidth=\dimexpr \marginparwidth + 2cm\relax
    \newcommand{\iey}[1]{\textcolor{dkdkgreen}{#1}}
    
    \newcommand{\gf}[1]{\textcolor{gfored}{#1}}
    \newcommand{\omi}[1]{\textcolor{red}{#1}}
    \newcommand{\omii}[1]{\textcolor{red}{#1}}

    \newcommand{\ieycomment}[1]{\todo[size=\scriptsize, linecolor=dkdkgreen, bordercolor=dkdkgreen, backgroundcolor=white]{\textcolor{dkdkgreen}{\textbf{@Ismail:} #1}}}

    \newcommand{\gfcomment}[1]{\todo[size=\scriptsize, linecolor=gfored, bordercolor=gfored, backgroundcolor=white]{\textcolor{gfored}{\textbf{@GF:} #1}}}

    \newcommand{\atbcomment}[1]{\todo[size=\scriptsize, linecolor=denim, bordercolor=denim, backgroundcolor=white]{\textcolor{denim}{\textbf{@atb:} #1}}}

    \newcommand{\omcomment}[1]{\todo[size=\scriptsize, linecolor=purple, bordercolor=purple, backgroundcolor=white]{\textcolor{purple}{\textbf{@OM:} #1}}}

\else
    
    \newcommand{\iey}[1]{#1}
    \newcommand{\ieycomment}[1]{}

    \newcommand{\atbcomment}[1]{}

    \newcommand{\gf}[1]{#1}
    \newcommand{\gfcomment}[1]{}

    \newcommand{\omi}[1]{#1}
    \newcommand{\omii}[1]{#1}
    \newcommand{\omcomment}[1]{}

\fi
\newcommand{\gfe}[1]{\textcolor{black}{#1}}
\newcommand{\om}[1]{\textcolor{black}{#1}}

\DeclareRobustCommand\circledtest[1]{\tikz[baseline=(char.base)]{
            \node[shape=circle,fill,inner sep=0pt] (char) {\scriptsize\textcolor{white}{#1}};}}

\usetikzlibrary{calc}
\DeclareRobustCommand\circledt[2]{\tikz[baseline=(char.base)]{
    \node[shape=circle, draw, fill=#1, inner sep=0pt] (char) {\scriptsize\textcolor{white}{#2}};}}

    \DeclareRobustCommand{\circlediii}[1]{\tikz[baseline=(char.base)]{\node[shape=circle,draw,inner sep=0pt,fill=white, text=black] (char) {\scriptsize#1};}}

        \DeclareRobustCommand{\circledv}[1]{\tikz[baseline=(char.base)]{\node[shape=circle,draw,inner sep=0pt,fill=white, text=black] (char) {\itshape#1};}}

\newcommand{\dingZero}{\circledtest{0}}
\newcommand{\dingOne}{\circledtest{1}}
\newcommand{\dingTwo}{\circledtest{2}}
\newcommand{\dingThree}{\circledtest{3}}
\newcommand{\dingFour}{\circledtest{4}}
\newcommand{\dingFive}{\circledtest{5}}
\newcommand{\dingSix}{\circledtest{6}}

\newcommand{\src}[0]{{\texttt{src}}}
\newcommand{\dst}[0]{{\texttt{dst}}}
\newcommand{\tras}[0]{$t_{RAS}$}
\newcommand{\trp}[0]{$t_{RP}$}
\newcommand{\act}[0]{\texttt{ACT}}
\newcommand{\pre}[0]{\texttt{PRE}}

\newcommand{\rref}[0]{\texttt{R$_{\texttt{REF}}$}}
\newcommand{\rcom}[0]{\texttt{R$_{\texttt{COM}}$}}
\newcommand{\vref}[0]{\texttt{V$_{\texttt{REF}}$}}
\newcommand{\vcom}[0]{\texttt{V$_{\texttt{COM}}$}}
\newcommand{\apaAnd}[0]{{\texttt{ACT}~\rref{}~$\rightarrow$~\texttt{PRE}~$\rightarrow$~\texttt{ACT}~\rcom{}}}

\newcommand{\apatrng}[0]{{\texttt{ACT}~\texttt{R0}~$\rightarrow$~\texttt{PRE}~$\rightarrow$~\texttt{ACT}~\texttt{R3}}}

\makeatletter
\g@addto@macro{\normalsize}{%
  \setlength{\abovedisplayskip}{0pt plus 1pt minus 1pt}
\setlength{\belowdisplayskip}{0pt plus 1pt minus 1pt}
 \setlength{\intextsep}{2pt plus 1pt minus 1pt}
 \setlength{\textfloatsep}{3pt plus 1pt minus 1pt}
  \setlength{\dbltextfloatsep}{13pt plus 1pt minus 1pt}
}
 \makeatother

\def\BibTeX{{\rm B\kern-.05em{\sc i\kern-.025em b}\kern-.08em
    T\kern-.1667em\lower.7ex\hbox{E}\kern-.125emX}}

\def\UrlBreaks{\do\/\do-\/\do.\/\do:}

\expandafter\def\expandafter\UrlBreaks\expandafter{\UrlBreaks
  \do\a\do\b\do\c\do\d\do\e\do\f\do\g\do\h\do\i\do\j
  \do\k\do\l\do\m\do\n\do\o\do\p\do\q\do\r\do\s\do\t
  \do\u\do\v\do\w\do\x\do\y\do\z\do\A\do\B\do\C\do\D
  \do\E\do\F\do\G\do\H\do\I\do\J\do\K\do\L\do\M\do\N
  \do\O\do\P\do\Q\do\R\do\S\do\T\do\U\do\V\do\W\do\X
  \do\Y\do\Z}
  
\newcommand{\squishlist}{
 \begin{list}{$\circ$}
  { \setlength{\itemsep}{0pt}
     \setlength{\parsep}{0pt}
     \setlength{\topsep}{0pt}
     \setlength{\partopsep}{0pt}
     \setlength{\leftmargin}{1em}
     \setlength{\labelwidth}{1em}
     \setlength{\labelsep}{0.5em} } }

\newcommand{\squishsublist}{
\begin{list}{$\rightarrow$}
 { \setlength{\itemsep}{0pt}
    \setlength{\parsep}{0pt}
    \setlength{\topsep}{-10em}
    \setlength{\partopsep}{-3pt}
    \setlength{\leftmargin}{1em}
    \setlength{\labelwidth}{1em}
    \setlength{\labelsep}{0.5em} } }

\newcommand{\squishend}{
  \end{list}  }

\newcommand{\li}{(\textit{i})}
\newcommand{\lii}{(\textit{ii})}

\newcommand\pnmref{\cite{Kautz69,stone1970logic,farmahini2015nda,babarinsa2015jafar,devaux2019true,ghiasi2022genstore,gomez2021benchmarkingcut,gomez2022benchmarking,syncron,singh2020nero,skhynixpim,ke2021near,giannoula2022sparsep,shin2018mcdram,cho2020mcdram,denzler2021casper,asghari2016chameleon,IRAM_Micro_1997,C_RAM_1999,CASES_MVX,Xi_2015,sun2021abc,matam2019graphssd,gokhale1995processing,hall1999mapping,MEMSYS_MVX,lockerman2020livia,ahn2015scalable,nai2017graphpim,boroumand2018google,lazypim, top-pim, gao2016hrl, kim2018grim, drumond2017mondrian, RVU, NIM, PEI, gao2017tetris,Kim2016,gu2016leveraging, boroumand2019conda, hsieh2016transparent, cali2020genasm, NDC_ISPASS_2014,pattnaik2016scheduling,akin2015data,hsieh2016accelerating,lee2015bssync,boroumand2021google,boroumand2022polynesia,amiraliphd,besta2021sisa,fernandez2020natsa,singh2019napel,kwon202125,lee2021hardware,niu2022184qps,Sparse_MM_LiM,azarkhish2016logic,azarkhish2018neurostream,guo20143d,akin2014hamlet,huang2020heterogeneous,dai2018graphh,liu2018processing,tsai:micro:2018:ams,gu2020ipim,DRAMA_CAL_2014,Asghari-Moghaddam_2016,huang2019active,kersey2017lightweight,li2019pims,zhuo2019graphq,zhang2018graphp,lim2017triple,smc_sim,HIVE,jang2019charon,IBM_ActiveCube,hadidi2017cairo,santos2018processing,lenjani2020fulcrum}\xspace}

\newcommand\pimref{\cite{farmahini2015nda,babarinsa2015jafar,devaux2019true,ghiasi2022genstore,gomez2021benchmarkingcut,gomez2022benchmarking,syncron,singh2020nero,skhynixpim,ke2021near,giannoula2022sparsep,shin2018mcdram,cho2020mcdram,denzler2021casper,asghari2016chameleon,IRAM_Micro_1997,C_RAM_1999,CASES_MVX,Xi_2015,sun2021abc,matam2019graphssd,gokhale1995processing,hall1999mapping,MEMSYS_MVX,lockerman2020livia,ahn2015scalable,nai2017graphpim,boroumand2018google,lazypim, top-pim, gao2016hrl, kim2018grim, drumond2017mondrian, RVU, NIM, PEI, gao2017tetris,Kim2016,gu2016leveraging, boroumand2019conda, hsieh2016transparent, cali2020genasm, NDC_ISPASS_2014,pattnaik2016scheduling,akin2015data,hsieh2016accelerating,lee2015bssync,boroumand2021google,boroumand2022polynesia,amiraliphd,besta2021sisa,fernandez2020natsa,singh2019napel,kwon202125,lee2021hardware,niu2022184qps,Sparse_MM_LiM,azarkhish2016logic,azarkhish2018neurostream,guo20143d,akin2014hamlet,huang2020heterogeneous,dai2018graphh,liu2018processing,tsai:micro:2018:ams,gu2020ipim,DRAMA_CAL_2014,Asghari-Moghaddam_2016,huang2019active,kersey2017lightweight,li2019pims,zhuo2019graphq,zhang2018graphp,lim2017triple,smc_sim,HIVE,jang2019charon,IBM_ActiveCube,hadidi2017cairo,santos2018processing,chi2016prime, Shafiee2016, seshadri2017ambit, seshadri2019dram, li2017drisa, seshadri2013rowclone, seshadri2016processing, deng2018dracc, xin2020elp2im, song2018graphr, song2017pipelayer,gao2019computedram, eckert2018neural, aga2017compute,dualitycache,seshadri2016buddy,seshadri.bookchapter17,seshadri2018rowclone,seshadri2015fast,li2016pinatubo,ferreira2022pluto,imani2019floatpim,he2020sparse,flashcosmos,truong2022adapting,truong2021racer,olgun2021quactrng,kim2019d,kim2018dram,bostanci2022dr,olgun2022pidram,ali2019memory,angizi2019graphide,li2018scope,subramaniyan2017parallel,zha2020hyper,fujiki2018memory,orosa2021codic,sharad2013ultra,rezaei2020nom,
gao2021parabit,choi2020flash,han2019novel,merrikh2017high,wang2018three,lue2019optimal,kim2021behemoth,wang2022memcore,han2021flash,kang2021s,lee2020neuromorphic,lee20223d,si2019dual,
simon2020blade,nag2019gencache,wang2019bit,al2020towards,kang2014energy,kim2021colonnade,jiang2020c3sram,jeloka201628,wang2023infinity,kang2015energy, imani2020dual,chang2016low, hajinazarsimdram,deng2019lacc,sutradhar2021look,sutradhar2020ppim,lenjani2020fulcrum,peng2023chopper,oliveira2022accelerating,singh2021fpga,oliveira2023dappa,oliveira2022methodologies,oliveira2022heterogeneous,shahroodi2023swordfish,chen2023simplepim,gupta2023evaluating,gomez2023evaluating,oliveira2023transpimlib,diab2023framework,mao2022genpip,singh2022accelerating}\xspace}

\newcommand\pumref{\cite{chi2016prime, Shafiee2016, seshadri2017ambit, seshadri2019dram, li2017drisa, seshadri2013rowclone, seshadri2016processing, deng2018dracc, xin2020elp2im, song2018graphr, song2017pipelayer,gao2019computedram, eckert2018neural, aga2017compute,dualitycache,besta2021sisa,seshadri2016buddy,seshadri.bookchapter17,seshadri2018rowclone,seshadri2015fast,li2016pinatubo,ferreira2022pluto,imani2019floatpim,he2020sparse,flashcosmos,truong2022adapting,truong2021racer,olgun2021quactrng,kim2019d,kim2018dram,bostanci2022dr,olgun2022pidram,ali2019memory,angizi2019graphide,li2018scope,subramaniyan2017parallel,zha2020hyper,fujiki2018memory,orosa2021codic,sharad2013ultra,rezaei2020nom,gao2021parabit,choi2020flash,han2019novel,merrikh2017high,wang2018three,lue2019optimal,kim2021behemoth,wang2022memcore,han2021flash,kang2021s,lee2020neuromorphic,lee20223d,si2019dual,
simon2020blade,nag2019gencache,wang2019bit,al2020towards,kang2014energy,kim2021colonnade,jiang2020c3sram,jeloka201628,wang2023infinity,kang2015energy,imani2020dual, chang2016low,hajinazarsimdram,deng2019lacc,sutradhar2021look,sutradhar2020ppim,peng2023chopper,shahroodi2023swordfish,fernandez2024matsa}\xspace}

\newcommand\drampim{\cite{ahn2015scalable,akin2014hamlet,akin2015data,ali2019memory,amiraliphd,angizi2019graphide,Asghari-Moghaddam_2016,asghari2016chameleon,azarkhish2016logic,azarkhish2018neurostream,babarinsa2015jafar,besta2021sisa,boroumand2018google, boroumand2019conda,boroumand2021google,boroumand2022polynesia,bostanci2022dr, cali2020genasm,CASES_MVX,chi2016prime,cho2020mcdram,C_RAM_1999,dai2018graphh,de2018design, deng2018dracc,devaux2019true,DRAMA_CAL_2014, drumond2017mondrian,farmahini2015nda,fernandez2020natsa,ferreira2022pluto,gao2016hrl,gao2017tetris,gao2019computedram,giannoula2022sparsep,gomez2021benchmarkingcut,gomez2022benchmarking,gu2020ipim,guo20143d,hsieh2016accelerating,hsieh2016transparent,huang2019active,huang2020heterogeneous,IRAM_Micro_1997,ke2021near,kersey2017lightweight,kim2018dram,kim2018grim,kim2019d,kwon202125,lazypim,lee2021hardware,li2017drisa,li2018scope,li2019pims,liu2018processing,nai2017graphpim, NDC_ISPASS_2014, NIM,niu2022184qps,olgun2021quactrng,olgun2022pidram,pattnaik2016scheduling, PEI, RVU,seshadri.bookchapter17, seshadri2013rowclone,seshadri2015fast,seshadri2016buddy, seshadri2016processing, seshadri2017ambit,seshadri2018rowclone, seshadri2019dram,shin2018mcdram,singh2019napel,singh2020nero,skhynixpim,Sparse_MM_LiM,sun2021abc,syncron,top-pim,tsai:micro:2018:ams, xin2020elp2im,Xi_2015,zhang2018graphp,zhuo2019graphq,lim2017triple,orosa2021codic,smc_sim,HIVE,jang2019charon,IBM_ActiveCube,rezaei2020nom,hall1999mapping,hadidi2017cairo,santos2018processing,MEMSYS_MVX}}

\newcommand\srampim{\cite{aga2017compute,denzler2021casper,dualitycache,eckert2018neural,gokhale1995processing,lockerman2020livia}\xspace}

\newcommand\ambit{\cite{seshadri2017ambit,seshadri2019dram,seshadri2015fast,seshadri.bookchapter17,seshadri2016buddy,seshadri2016processing}\xspace}

\newcommand{\dataMovement}[0]{\cite{mutlu2013memory,mutlu2015research,dean2013tail,kanev_isca2015,ferdman2012clearing,wang2014bigdatabench,mutlu2019enabling,mutlu2019processing,mutlu2020intelligent,ghose.ibmjrd19,mutlu2020modern,oliveira2021damov,boroumand2018google,boroumand2021google,wang2016reducing, pandiyan2014quantifying,koppula2019eden,kang2014co,mckee2004reflections,wilkes2001memory,kim2012case,wulf1995hitting,ghose.sigmetrics20,ahn2015scalable,PEI,hsieh2016transparent,wang2020figaro,sites1996}}

\newcommand{\ignore}[1]{}

\renewcommand{\thesection}{\textbf{\Roman{section}}}

\begin{document}
\vspace{-10ex}
\title{\fontsize{20}{20}\selectfont{Memory-Centric Computing: \\ Recent Advances in Processing-in-DRAM}\vspace{-0.5ex}}
\author{\fontsize{11}{11}\selectfont Onur Mutlu, Ataberk Olgun, Geraldo F. Oliveira, Ismail E. Yuksel
\\
\fontsize{10}{12}\selectfont 
ETH Z{\"u}rich \vspace{-3ex}
}

\setstretch{0.85}

\maketitle
\thispagestyle{plain}

\pagestyle{plain}

\noindent \textbf{\textit{Abstract}}---{\em Memory-centric computing} aims to enable computation capability in and near all places where data is generated and stored. As such, it can greatly reduce the large negative performance and energy impact of data access and data movement, by 1) fundamentally avoiding data movement, 2) reducing data access latency \& energy, and 3) exploiting large parallelism of memory arrays. Many recent studies show that memory-centric computing can largely improve system performance \& energy efficiency. Major industrial vendors and startup companies have recently introduced memory chips with sophisticated computation capabilities. 
Going forward, both hardware and software stack should be revisited and designed carefully to take advantage of memory-centric computing. 

This work describes several major recent advances in memory-centric computing, specifically in {\em Processing-in-DRAM}, a paradigm where the operational characteristics of a DRAM chip are exploited and enhanced to perform computation on data stored in DRAM. Specifically, we describe 
1) new techniques that slightly modify DRAM chips to enable both enhanced computation capability and easier programmability, 
2) new experimental studies that demonstrate the functionally-complete bulk-bitwise computational capability of real commercial off-the-shelf DRAM chips, without any modifications to the DRAM chip or the interface, and 
3) new DRAM designs that improve access granularity \& efficiency, unleashing the true potential of Processing-in-DRAM. 
\vspace{-0.5ex}

\section{Memory-Centric Computing}

\emph{Data movement} between computation units (e.g., CPUs, GPUs, ASICs) and main memory (e.g., DRAM) is a major \emph{performance} and \emph{energy bottleneck} in current processor-centric computing systems~\dataMovement{}, and is expected to worsen due to the increasing data intensiveness of modern applications, e.g., machine learning~\cite{brown2020language,devlin2019bert,boroumand2021google,oliveira2022accelerating,heo2024neupims,zhou2022transpim,park2024attacc,seo2024ianus, rhyner2024pim,yun2024duplex,jang2024smart} and genomics~\cite{alser2020accelerating,singh2021fpga,alser2022from,kim2018grim,ghiasi2022genstore,ghiasi2024megis,cali2020genasm,cali2022segram,senolcali.bib2019}.
To mitigate the overheads caused by data movement, various works propose \gls{pim} architectures~\pimref.
There are two main approaches to \gls{PIM}~\cite{ghose.ibmjrd19, mutlu2020modern, mutlu2019processing}:
\li~\gls{pnm}~\pnmref, where computation logic is added near the memory arrays (e.g., in a DRAM chip, next to each bank, or at the logic layer of a 3D-stacked memory~\cite{HMC2, HBM, lee2016simultaneous, hbm2, hbm3, kim2024present, ahn2015scalable,PEI,hsieh2016transparent,boroumand2018google,boroumand2022polynesia,boroumand2021google,boroumand2019conda,singh2021fpga,oliveira2022accelerating}); and \lii~\gls{pum}~\pumref, where computation is performed by exploiting the analog operational properties of the memory circuitry.

Both approaches, offering different tradeoffs, are important to exploit the full potential of PiM.
\gls{pum} has two major advantages over \gls{pnm}: 1) \gls{pum} \emph{fundamentally} reduces data movement by performing computation \emph{in situ}, while data movement still occurs between computation units and memory arrays in \gls{pnm}; 
2) \gls{pum} exploits the large internal bandwidth and parallelism available \emph{inside} the memory arrays, while \gls{pnm} is bottlenecked by the memory's internal data buses. In contrast, \gls{pnm} can enable a wider set of functions (including complete processors) to be more easily implemented and exploited near memory due to its use of conventional logic. 

\gls{pim} (both~\gls{pum} and~\gls{pnm}) can be implemented in (i.e., using or near) different memory technologies~\cite{ghose.ibmjrd19, mutlu2020modern, mutlu2019processing}, including SRAM (e.g.,\srampim), DRAM (e.g.,\drampim), NAND flash (e.g,\cite{flashcosmos,gao2021parabit,choi2020flash,han2019novel,merrikh2017high,wang2018three,lue2019optimal,kim2021behemoth,wang2022memcore,han2021flash,kang2021s,lee2020neuromorphic,lee20223d,ghiasi2022genstore,jun2015bluedbm,ghiasi2024megis}), or emerging (e.g., \cite{fujiki2018memory,imani2019floatpim,Kim2016,li2016pinatubo,Shafiee2016,song2017pipelayer,song2018graphr,truong2021racer,zha2020hyper,truong2022adapting,sharad2013ultra,shahroodi2023swordfish}).
We focus on DRAM~\cite{dennard1968field} due to its dominance as the main memory technology and very large capacity that can house many data-intensive workloads at reasonably low access latency.

\section{Processing-in-DRAM}

Many works demonstrate \gls{pnd} and \gls{pud}.

\gls{pnd} architectures are designed to accelerate important applications by adding specialized or general-purpose compute units next to DRAM banks or arrays, either inside the DRAM chip or inside the logic layer of 3D-stacked DRAM architectures. Prominent examples include acceleration of graph analytics~\cite{ahn2015scalable,PEI,nai2017graphpim,besta2021sisa}, machine learning~\cite{brown2020language,devlin2019bert,boroumand2021google,oliveira2022accelerating,gomez2023evaluating,rhyner2024pim,gogineni2024swiftrl,heo2024neupims,zhou2022transpim,park2024attacc,seo2024ianus,yun2024duplex},
 mobile workloads~\cite{boroumand2018google}, genome analytics~\cite{kim2018grim,cali2020genasm,cali2022segram,diab2023framework}, databases~\cite{boroumand2019conda,amiraliphd,lazypim,boroumand2022polynesia,drumond2017mondrian,hsieh2016accelerating,RVU},
climate modeling~\cite{singh2020nero,singh2022accelerating},
time series analysis~\cite{fernandez2020natsa}, 
security functions~\cite{gu2016leveraging,xiong2022secndp,gupta2023evaluating,aga2017invisimem}, 
data manipulation~\cite{akin2015data,akin2014hamlet,boroumand2018google}, 
and GPU workloads~\cite{hsieh2016transparent,pattnaik2016scheduling}.

\begin{figure}[!b]
    \centering
    \includegraphics[width=1\linewidth]{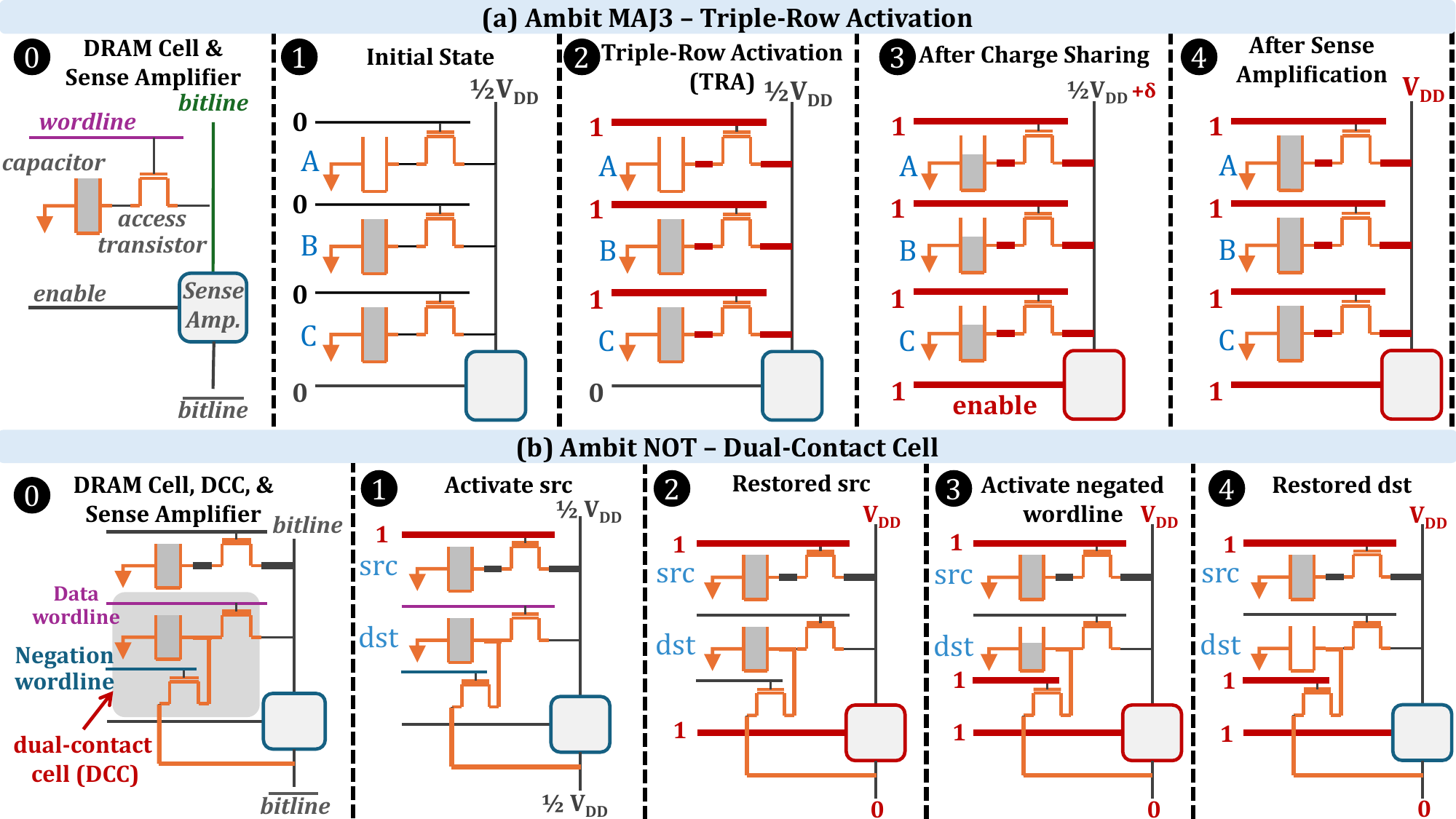}
    \caption{\textbf{An example of performing the MAJority-of-three operation (i.e., \texttt{MAJ3(A,B,C)}) (a) and the NOT operation (i.e., \texttt{dst=NOT(src)}) in Ambit~\cite{seshadri2017ambit}}. 
    In (a), we focus on DRAM cell and sense amplifier operations (\dingZero{}). 
    Initially, cells A, B, C, and bitline have voltage levels of GND, VDD, VDD, and VDD/2, respectively (\dingOne{}). 
    We first perform a triple-row activation (TRA) to simultaneously activate cells A, B, and C (\dingTwo{}). When the wordlines of all three cells are raised simultaneously, charge sharing results in a positive deviation on the bitline because at least two of the cells are charged (\dingThree{}). Therefore, after sense amplification, the sense amplifier drives the bitline to VDD, which then fully charges all three cells (\dingFour{}). The final state of the bitline is, thus, the MAJority function of the charged state of the three cells A, B, and C. If one of the cells (say C) is set to GND (VDD), the final state would be the AND (OR) of the other two (A and B). To simplify the explanation, we assume no process variation and noise, but the Ambit paper and later works~\cite{hajinazarsimdram,seshadri2017ambit} evaluate these effects.
    \\Ambit-NOT (b) introduces the dual-contact cell (DCC), which is a DRAM cell with two transistors. In a DCC, one transistor connects the cell capacitor to the bitline\gf{,} i.e., data wordline, and the other transistor connects the cell capacitor to the bitline-bar, i.e., negation wordline (\dingZero{}).
    Initially, \src{} and \dst{} cells each have a voltage level of VDD, and bitline and bitline-bar are precharged to VDD/2. To perform the NOT operation, we first activate the \src{} cell (\dingOne{}). The activation drives the bitline to the value corresponding to the \src{}, VDD in this case, and the bitline-bar to the negated value, i.e., GND (\dingTwo{}). Second, Ambit activates the negation wordline. Doing so enables the transistor that connects the DCC to the bitline-bar. This results in the bitline-bar sharing its charge with the \dst{} cell (\dingThree{}). Since the bitline-bar is already at a stable voltage level of GND, it overwrites the value in the DCC capacitor with GND, th\omi{ereby} copying the negated value of the \src{} cell into the \dst{} cell (\dingFour{}).
    \\The original Ambit paper (see Section 5 in~\cite{seshadri2017ambit}) proposes a DRAM subarray design that makes the implementation of triple-row activation low overhead by restricting TRA to an isolated set of DRAM rows that can be used for computation. It also describes circuit-level issues \& system and programming support needed for Ambit and evaluates the hardware cost of modifications made to the DRAM chip and the memory controller. A later work~\cite{seshadri2019dram} describes some outstanding issues in Ambit-like bulk-bitwise Processing-using-DRAM substrates. 
    }
    \label{fig:ambit_ops}
\end{figure}

\begin{figure}[!hb]
\centering
\begin{subfigure}[b]{1\linewidth}
\centering
\includegraphics[width=\linewidth]{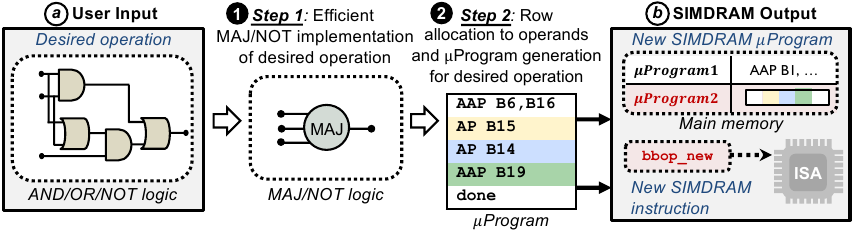}
\label{subfig:simdram_step1_2}
\end{subfigure}%
\\
\vspace{-1.2em}
\begin{subfigure}[b]{1\linewidth}
\centering
\includegraphics[width=\linewidth]{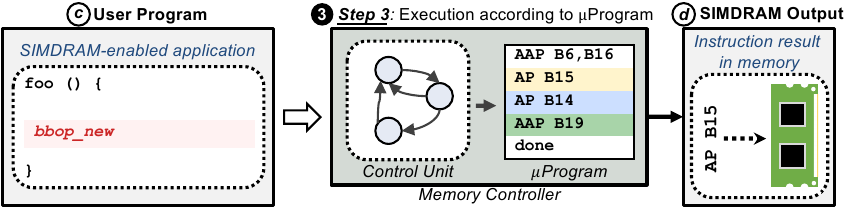}
\label{subfig:simdram_step2_3}
\end{subfigure}%
\caption{\textbf{Overview of the SIMDRAM framework~\cite{hajinazarsimdram}.} SIMDRAM consists of three key
steps to enable a user-specified desired operation in DRAM: 
1)~building an efficient MAJ/NOT-based representation of the desired operation, 
2)~mapping the operation input and output operands to DRAM rows and to the required DRAM commands that produce the desired operation, and 
3)~executing the operation. 
The first two steps give users the flexibility to implement and compute any desired operation in DRAM efficiently. 
The goal of the first step is to use logic optimization to minimize the number of DRAM row activations and, \omi{thus}, the computation latency required to perform a specific operation. Accordingly, the first step (\dingOne{}) takes as input the AND/OR/NOT-based implementation of the designed operation (labeled \circlediii{a} in the figure) and derives the operation's optimized MAJ/NOT-based implementation (i.e., the optimized majority inverter graph).
The second step (\dingTwo{}) translates the optimized MAJ/NOT-based implementation into DRAM row activations,
i.e., Ambit TRA~\cite{seshadri2017ambit} and RowClone~\cite{seshadri2013rowclone} operations. This step includes 
1)~mapping the operands to the designated rows in DRAM and 
2)~defining the sequence of DRAM row activations required to perform the computation associated with the optimized MAJ/NOT implementation. SIMDRAM chooses the operand-to-row mapping and the sequence of DRAM row activations to minimize the number of DRAM row activations required for a specific operation. The output of the second step \circlediii{b} is stored as a microprogram ($\mu$Program) in the memory controller, associated with the desired operation, \emph{bbop$\mathunderscore$new}.
The third step (\dingThree{}) is to program the memory controller to issue the sequence of DRAM row activations to the appropriate rows in DRAM to perform the computation of the operation from start to end. 
When the user program (\circlediii{c}) encounters a SIMDRAM instruction (called \emph{bbop$\mathunderscore$new}), the instruction is shipped to the memory controller\omi{,} which invokes the associated $\mu$Program and executes the operation as specified by the $\mu$Program.
To this end, SIMDRAM uses a control unit in the memory controller that transparently executes the sequence of DRAM row activations for each specific PuD operation executed by a user program.
Once the $\mu$Program is complete, the result of the
operation \circlediii{d} is held in DRAM.
Figure adapted from~\cite{hajinazarsimdram}.
}
\label{fig:simdram_framework}
\end{figure}

\gls{pud} architectures use DRAM to perform primitive operations (e.g., data copy, initialization, bitwise operations), on top of which different applications and software stacks can be built. We highlight a few major works we build on. RowClone~\cite{seshadri2013rowclone} demonstrates that data copy and initialization can be accelerated within a DRAM subarray by performing two consecutive \omi{row} activations, which lead to data in the first activated row to be copied to the second activated row via the sense amplifiers that are connected to both rows.\footnote{This work also demonstrates that DRAM operational principles can be used to copy data between different banks and subarrays, which later work~\cite{chang2016low,rezaei2020nom,wang2020figaro} improves by adding further logic to facilitate.}  Ambit~\cite{seshadri2015fast,seshadri2017ambit,seshadri2019dram} demonstrates that 
i)~concurrently activating three DRAM rows leads to the computation of the bitwise MAJority function (and thus the AND and OR functions) on the contents of the three rows due to the charge sharing principles that govern the operation of the shared bitlines and sense amplifiers (Fig.~\ref{fig:ambit_ops}a), 
ii)~bitwise NOT of a row can be performed through the sense amplifier, with modifications to DRAM circuitry (Fig.~\ref{fig:ambit_ops}b). Ambit provides a DRAM chip architecture that can exploit such triple-row activation (TRA), NOT, and RowClone operations. SIMDRAM~\cite{hajinazarsimdram} shows that, via a new software/hardware cooperative framework (Fig.~\ref{fig:simdram_framework}), \emph{any} 
operation (e.g., multiplication, division, convolution) that can be expressed as a logic circuit consisting of AND, OR, NOT gates can be implemented and seamlessly programmed using the Ambit substrate.  

Many operations envisioned by these \gls{pud} works can \emph{already} be performed in \emph{real unmodified} commercial off-the-shelf (COTS) DRAM chips, by violating manufacturer-recommended DRAM timing parameters. Recent works show that COTS DRAM chips can perform 1) data copy \& initialization~\cite{gao2019computedram,olgun2022pidram} (as in RowClone~\cite{seshadri2013rowclone}), 2) three-input bitwise MAJ and two-input AND \& OR operations~\cite{gao2019computedram,gao2022frac,olgun2023dram} (as in Ambit~\ambit),
and 3) true random number generation \& physical unclonable functions~\cite{kim2019d, olgun2021quactrng,kim2018dram}.

\section{Challenges \& Overview}

To realize the full potential and benefits of \gls{pid}, and more generally \gls{pim}, a number of important challenges need to be solved~\cite{mutlu2020modern, ghose.ibmjrd19,mutlu2019processing}.
This work tackles several of these major challenges.

First, enabling widespread use of PiM on a wide variety of important workloads requires \gls{pim} systems to i) be easy to program and seamlessly compile workloads into~\cite{mutlu2020modern, ghose.ibmjrd19,mutlu2019processing} and ii) support a wide range of computation primitives and capabilities. Second, it is important to demonstrate the potential feasibility and capabilities of future \gls{pim} architectures, especially \gls{pum} ideas that exploit analog operational capabilities of memory chips, ideally on real hardware. Third, it is important to design the memory (e.g., DRAM) chip architectures to efficiently support processing capability in a programmable manner.

We highlight several recent works~\cite{oliveira2024mimdram,yuksel2024functionallycomplete,yuksel2024simultaneous,olgun2024sectored,yuksel2023pulsar} that tackle these challenges for \gls{pid} systems: 1) MIMDRAM~\cite{oliveira2024mimdram}, for enabling easier programmability and enhanced computation capability, 2) new experimental studies using commercial off-the-shelf (COTS) DRAM chips~\cite{yuksel2024functionallycomplete,yuksel2024simultaneous,yuksel2023pulsar}, which demonstrate previously-unknown computational capabilities of real unmodified DRAM chips, and 3) Sectored DRAM~\cite{olgun2024sectored}, a new fine-grained DRAM design, that enables an efficient and easier-to-program DRAM substrate for \gls{pim}.

\section{MIMDRAM}
\begin{figure}[!htp]
    \centering
    \includegraphics[width=\linewidth]{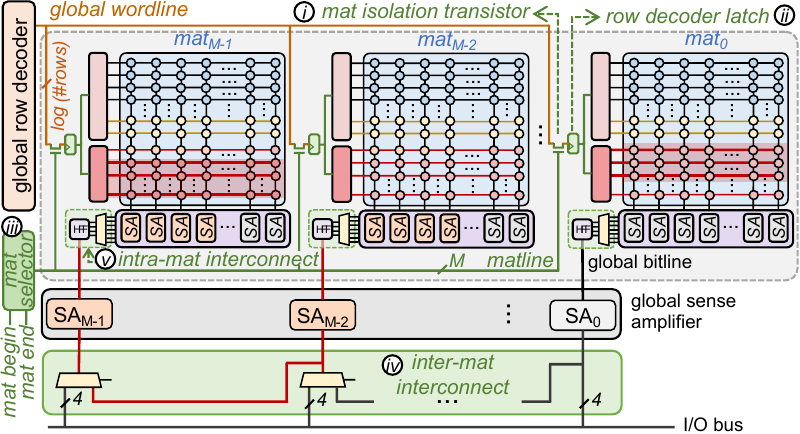}
    \caption{\textbf{Overview of the DRAM subarray and bank organization of MIMDRAM~\cite{oliveira2024mimdram}.} Green-colored boxes represent newly added or modified hardware components.
    To enable fine-grained PuD execution, MIMDRAM modifies Ambit's subarray and the DRAM bank with three new hardware structures: 
    the \emph{mat isolation transistor} (\circledv{i}), the \emph{row decoder latch} (\circledv{ii}), and the \emph{mat selector} (\circledv{iii}). 
    At a high level, the \emph{mat isolation transistor} allows for the independent access and operation of \omi{each} DRAM mat within a subarray while 
    the \emph{row decoder latch} enables the execution of a PuD operation in a range of DRAM mats that the \emph{mat selector} defines.
    MIMDRAM implements an \emph{inter-mat interconnect} (\circledv{iv}) to enable data movement across different mats by slightly modifying the connection between the I/O bus and the global sense amplifier. MIMDRAM adds a 2:1 multiplexer to each set of four 1-bit sense amplifiers in the global sense amplifier, selecting whether the data written to the sense amplifier set ($SA_{i}$) comes from the I/O bus or the neighboring sense amplifier set ($SA_{i-1}$). MIMDRAM enables data movement across columns within a DRAM mat through an \emph{intra-mat interconnect} (\circledv{v}), which works by modifying the sequence of steps in the column access operation (hence without any hardware modification to the DRAM subarray structure). 
    The intra-mat interconnect leverages the fact that
    1)~local bitlines in a mat already share an interconnection link via the helper flip-flops (HFFs) and 
    2)~these HFFs can latch and amplify the local row buffer's data.
    Figure adapted from~\cite{oliveira2024mimdram}.
    }
    \label{fig:mimdram_overview}
\end{figure}

\begin{figure}[!hbp]
    \centering
    \includegraphics[width=\linewidth]{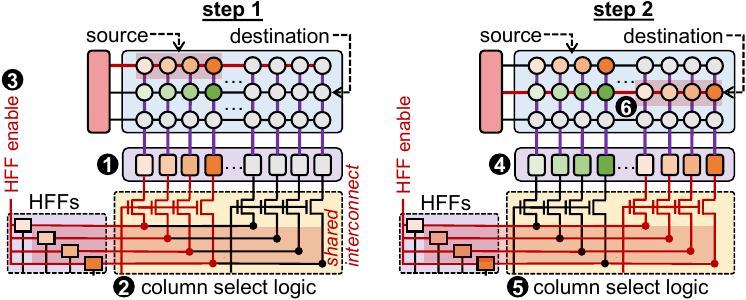}
    \caption{\textbf{Intra-mat data movement in MIMDRAM~\cite{oliveira2024mimdram}.}~{To enable data movement across columns \emph{within} a DRAM mat, MIMDRAM implements an intra-mat interconnect (\gfe{\circledv{v}} in Fig.\ref{fig:mimdram_overview}), which does not require any hardware \gfe{modification}. Instead, \gfe{MIMDRAM} modifies the sequence of steps DRAM executes during a column access \gfe{to realize an intra-mat data movement operation}. 
    There are two key observations that enable the intra-mat interconnect. First, we observe that the local bitlines of a DRAM mat already share an interconnection path via the HFFs and column select logic (as this figure illustrates). Second, the HFFs in a DRAM mat can latch and amplify the local row buffer’s data.}
    \\To manage intra-mat data movement, MIMDRAM exposes a new DRAM
    command to the memory controller called \texttt{LC-MOV} (\underline{l}o\underline{c}al I/O
    \underline{mo}ve). {The \texttt{LC-MOV} command takes as input: (i) the logical mat range \textit{[mat begin,mat end]} of the target row, (ii) the row and column addresses of the \emph{source} DRAM row and column; and (iii) the row and column addresses of the \emph{destination} DRAM row and column.  With the intra-mat interconnect and new DRAM command, MIMDRAM can move four bits of data from a source row and column (\texttt{row$_{src}$}, \texttt{column$_{src}$}) to a destination row and column (\texttt{row$_{dst}$}, \texttt{column$_{dst}$}) in the same mat ($mat_{M}$).}
    \gfe{An \texttt{LC-MOV} command is \emph{transpa\om{r}ently} generated by MIMDRAM control unit based on the source and destination mat addresses in a \texttt{bbop\_mov} instruction (which MIMDRAM compiler generates\om{; see~\cite{oliveira2024mimdram}}): if the source and destination mats\om{'} addresses are \emph{the same}, MIMDRAM control unit translates the data movement instruction into an \texttt{LC-MOV} command; otherwise, \om{into} a 
    \texttt{GB-MOV} command~\om{(\underline{g}lo\underline{b}al I/O
    \underline{mo}ve; see~\cite{oliveira2024mimdram} and Fig.\ref{fig:vector_reduction})}.}
    \\Once the memory controller receives an \texttt{LC-MOV} command, it performs two steps. In the \emph{first step}, the memory controller performs an \texttt{ACT}--\texttt{RD}--\texttt{PRE} targeting 
    \texttt{row$_{src}$}, \texttt{column$_{src}$} in $mat_{M}$. The \texttt{ACT} loads \texttt{row$_{src}$} to 
    $mat_{M}$'s local sense amplifier (\dingOne{}). The \texttt{RD} moves four bits from  \texttt{row$_{src}$}, 
    as indexed by \texttt{column$_{src}$}, into the mat's helper flip-flops (HFFs) by enabling the appropriate 
    transistors in the column select logic~(\dingTwo{}). The HFFs are then enabled by transitioning the 
    \emph{HFF enable} signal from low to high. This allows the HFF to \emph{latch} and \emph{amplify} the 
    selected four-bit data column from the local sense amplifier~(\dingThree{}). The \texttt{PRE} closes 
    \texttt{row$_{src}$}. Until here, the \texttt{LC-MOV} command operates exactly as a regular  \texttt{ACT}--
    \texttt{RD}--\texttt{PRE} command sequence. However, differently from a regular \texttt{ACT}--\texttt{RD}--
    \texttt{PRE}, the \texttt{LC-MOV} command does \emph{not} lower the \emph{HFF enable} signal when the 
    \texttt{RD} finishes. This allows the four-bit data from \texttt{column$_{src}$} to reside in the mat's HFF. 
    In the \emph{second step}, the memory controller performs an \texttt{ACT}--\texttt{WR}--\texttt{PRE} 
    targeting \texttt{row$_{dst}$}, \texttt{column$_{dst}$} in $mat_{M}$. The \texttt{ACT} loads 
    \texttt{row$_{dst}$} into the mat's local row buffer (\dingFour{}), and the \texttt{WR} asserts the column 
    select logic to \texttt{column$_{dst}$}, creating a path between the HFF and the local row buffer 
    (\dingFive{}). Since the \emph{HFF enable} signal is kept high, the HFFs {do} \emph{not} sense and latch 
    the data from \texttt{column$_{dst}$}. Instead, the HFFs overwrite the data stored in the local sense 
    amplifier with the previously four-bit data latched from \texttt{column$_{src}$}. The new data stored in 
    the mat's local sense amplifier propagates through the local bitlines and is written to the destination 
    DRAM cells (\dingSix{}). {Figure adapted from~\cite{oliveira2024mimdram}.} 
    }
  
    \label{fig:intra_mat_network}
\end{figure}

\begin{figure}[!htbp]
    \centering
    \includegraphics[width=0.85\linewidth]{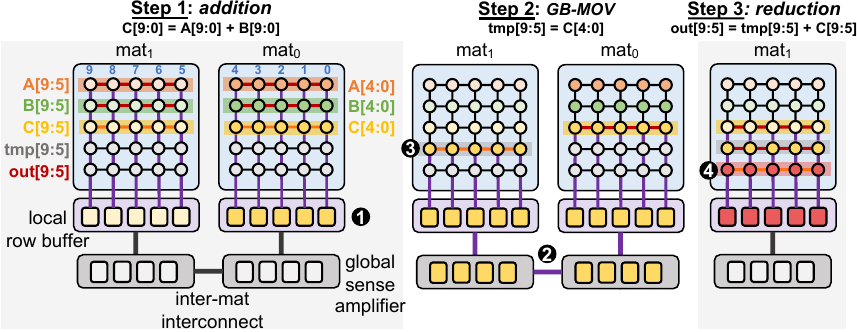}
    \caption{\textbf{An example of a PuD vector reduction, i.e.,~\texttt{out+=(A[i]+B[i])}, in MIMDRAM~\cite{oliveira2024mimdram}.} For illustration purposes, we assume that DRAM has only two mats, and the \omi{10 1-bit} data elements of the input arrays \texttt{A} and \texttt{B} are evenly distributed across the two DRAM mats. MIMDRAM executes a vector reduction in three steps.
    In the first step, MIMDRAM executes a PuD addition operation over the data in the two DRAM mats (\dingOne{}), storing the temporary output data \texttt{C} into the same mats where the computation takes place (i.e., \omi{\texttt{C} = \texttt{\{}\texttt{C[9:5]}$_{mat1}$, \texttt{\omii{C[4:0]}}$_{mat0}$\texttt{\}}}). 
    In the second step, MIMDRAM issues a \texttt{GB-MOV} (\om{global I/O move;} a new DRAM command to perform inter-mat data movement) to move part of the temporary output \omi{\texttt{C[4:0]}} stored in $mat_{0}$ to a temporary row \texttt{tmp} in $mat_{1}$ (\omi{\texttt{tmp[9:5]}$_{mat1}$$\leftarrow$\texttt{C[4:0]}$_{mat0}$}) via the inter-mat interconnect (\dingTwo{}--\dingThree{}), four bits, i.e., four data elements, at a time, which corresponds to the size of the helper flip-flops \omi{(not shown in the figure)}. MIMDRAM \emph{iteratively} executes step 2 until \emph{all} data elements of \omi{\texttt{C[4:0]}} are copied to $mat_{1}$.
    In the third step, once the \texttt{GB-MOV} finishes, MIMDRAM executes the final addition operation, i.e., \omi{\texttt{tmp[9:5]}~+~\texttt{C[9:5]}}, in $mat_{1}$. The final output of the vector reduction operation is stored in the destination row \texttt{out} in $mat_{1}$ (\dingFour{}). 
    Figure adapted from~\cite{oliveira2024mimdram}.
    }
    \label{fig:vector_reduction}
\end{figure}

\vspace{-5pt}
MIMDRAM is a hardware/software co-designed \gls{pud} system that introduces new mechanisms to allocate and control only the necessary resources for a given \gls{pud} operation. Major key ideas of MIMDRAM are 1) to leverage fine-grained DRAM (i.e., the ability to independently access smaller segments of a large DRAM row; see Section~\ref{sec:sectored-dram}) for \gls{pud} computation, 2) enable near-subarray reduction computation logic across DRAM mats, and 3) provide compiler support to transparently map vector operations to DRAM mats and subarrays. MIMDRAM provides a multiple-instruction multiple-data (MIMD) execution model~\cite{flynn1966very} in each DRAM subarray~\cite{kim2012case}, enabling different DRAM mats within a subarray to execute different \gls{pud} instructions (where each \gls{pud} instruction specifies bit-serial SIMD execution within one or more DRAM mats). MIMDRAM eases programmability and compilation by reducing the minimum granularity of \gls{pud} operations to bit-widths commonly targeted by modern vectorizing compilers and enabling flexibility in computation granularity.~Fig.~\ref{fig:mimdram_overview} presents an overview of MIMDRAM.
Fig.~\ref{fig:intra_mat_network} presents {an overview of} the intra-mat interconnect {in} MIMDRAM. 
Fig.~\ref{fig:vector_reduction} shows an example \gls{pud} vector reduction operation using MIMDRAM.

\begin{figure}[!htbp]
    \centering
    \includegraphics[width=0.85\linewidth]{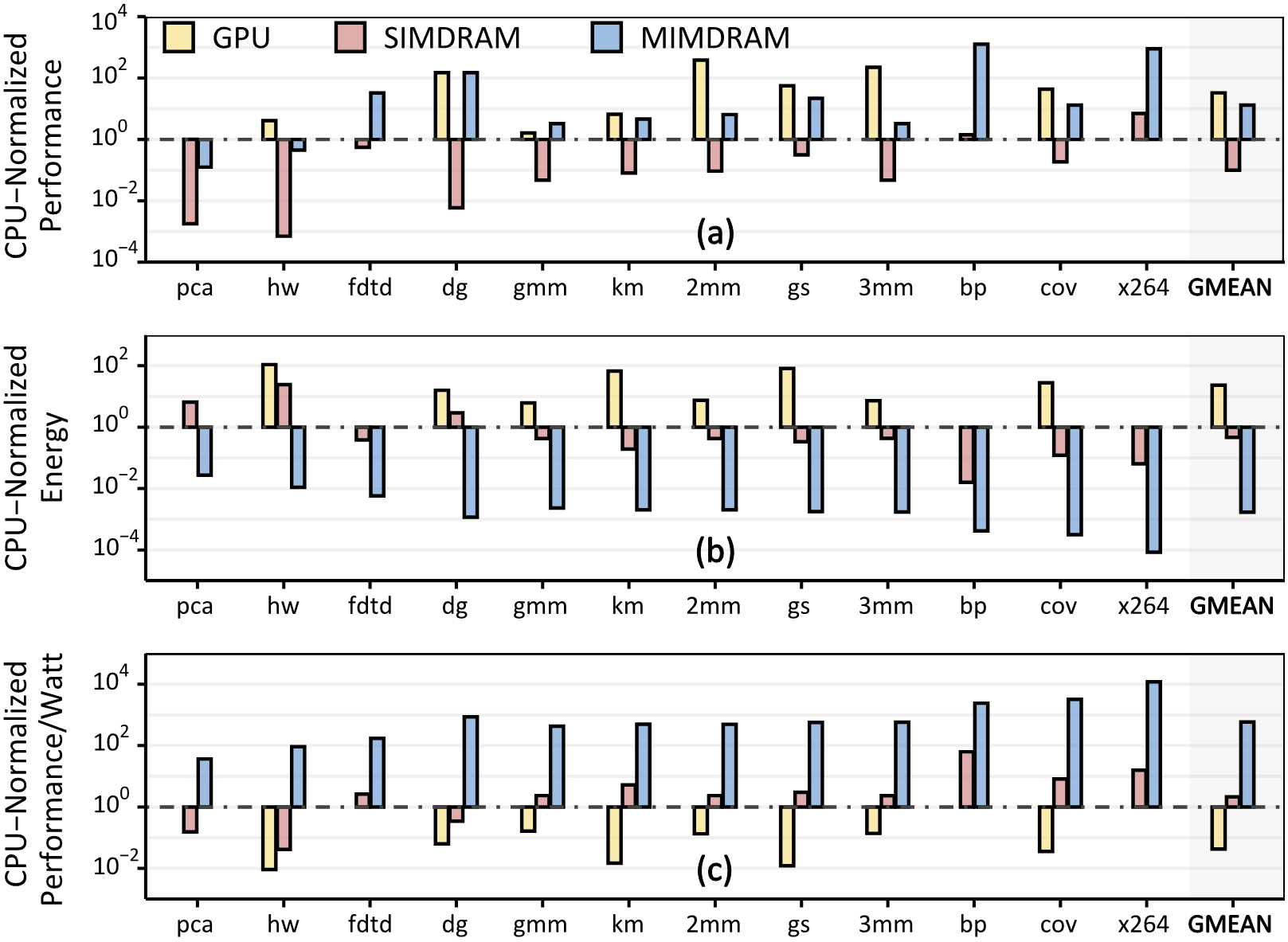}
    \caption{\textbf{CPU-normalized \gf{performance} (a), energy (b), and \omi{energy efficiency (}performance/Watt\omi{)} (c) results for processor-centric (i.e., Intel Skylake CPU~\cite{intelskylake} and NVIDIA A100 GPU~\cite{a100}) and memory-centric (i.e., SIMDRAM~\cite{hajinazarsimdram} and MIMDRAM~\cite{oliveira2024mimdram}) architectures executing 12 real-world applications.} \iey{We implement MIMDRAM and SIMDRAM using gem5 and evaluate their energy consumption using CACTI. We analyze 117 applications from SPEC 2017, SPEC 2006, Parboil, Phoenix, Polybench, Rodinia, and SPLASH-2 benchmark suites and identify 12 memory-bound, multi-threaded CPU applications where the most time-consuming loop can be auto-vectorized. These applications span various domains, including video compression, data mining, pattern recognition, medical imaging, and stencil computation.} \gf{For this analysis, we allow SIMDRAM and MIMDRAM to fully leverage bank and subarray-level parallelism in a DRAM rank by allowing in-DRAM operations to happen \omi{simultaneously} across \omii{all} 16 banks and 64 subarrays \omi{per bank}.} 
    MIMDRAM \omii{provides} (1)~13.2$\times$/0.22$\times$/173$\times$ the performance,  (2)~0.0017$\times$/0.00007$\times$/0.004$\times$ the energy consumption, and (3)~582.4$\times$/13612$\times$/272$\times$ the performance per Watt of the CPU/GPU/SIMDRAM baseline.
    \gf{In our analysis, we observe that MIMDRAM's end-to-end performance gains are limited by the throughput of the inter- and intra-mat interconnects, which are utilized during in-DRAM reduction operations. \omi{If we} consider only MIMDRAM's arithmetic throughput \omi{(i.e., no reduction operations)}, we observe that MIMDRAM \omii{provides} 272$\times$ and 11$\times$ the performance of the CPU and GPU baselines, respectively. We believe that combining PuD and PnD holistically, where auxiliary logic placed within the logic layer of 3D-stacked memories is used for high-throughput in-DRAM reduction and DRAM cells are used for high-throughput in-DRAM bulk arithmetic, would be beneficial to improve MIMDRAM's end-to-end performance.}
    We conclude that MIMDRAM \omi{enables effective exploitation of DRAM} bank\omi{-level} and subarray-level parallelism \omi{for massively-parallel bulk bitwise execution.}
    }
    \label{fig:mimdram_perf_energy}
\end{figure}

\begin{figure}[!htbp]
    \centering
    \includegraphics[width=0.85\linewidth]{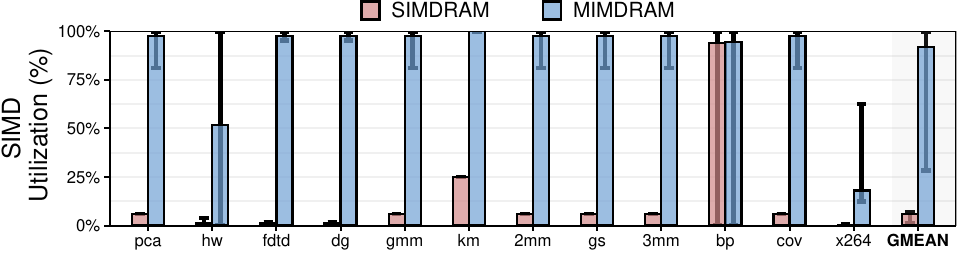}
    \caption{\textbf{SIMD utilization {(i.e., the fraction of SIMD lanes executing a useful operation)} of SIMDRAM and MIMDRAM for twelve real-world applications.} Whiskers extend to the minimum and maximum observed data point values. On average, across all twelve real-world applications, MIMDRAM provides 15.6$\times$ the SIMD utilization of SIMDRAM. This is because MIMDRAM matches the available SIMD parallelism in an application with the underlying PuD resources (i.e., PuD SIMD lanes) by using only as many DRAM mats as the maximum vectorization factor of a given application’s loop. In contrast, SIMDRAM always occupies all available PuD SIMD lanes (i.e., entire subarrays) for a given operation, resulting in low SIMD utilization for applications without a very-wide vectorization factor. We conclude that MIMDRAM greatly improves overall SIMD utilization for many applications.
    }
    \label{fig:mimdram_utilization}
\end{figure}

We evaluate MIMDRAM using twelve real-world applications and 495 multi-programmed application mixes. When using 64 DRAM subarrays per bank and 16 banks for PuD computation in a DRAM chip, MIMDRAM provides 1) 13.2×/0.22×/173× the performance, 2) 0.0017×/0.00007×/0.004× the energy consumption, 3) 582.4×/13612×/272× the performance per Watt of the CPU~\cite{intelskylake}/GPU~\cite{a100}/SIMDRAM~\cite{hajinazarsimdram} baseline (Fig.~\ref{fig:mimdram_perf_energy}) and 4) when using a single DRAM subarray, 15.6$\times$ the SIMD utilization, i.e., SIMD efficiency (Fig.~\ref{fig:mimdram_utilization}) of the prior state-of-the-art \gls{pud} framework, SIMDRAM.

\vspace{-15pt}

\vspace{-15pt}
\newpage
\section{Capabilities of Real COTS DRAM Chips}

\begin{figure}[!htbp]
    \centering
    \includegraphics[width=1\linewidth]{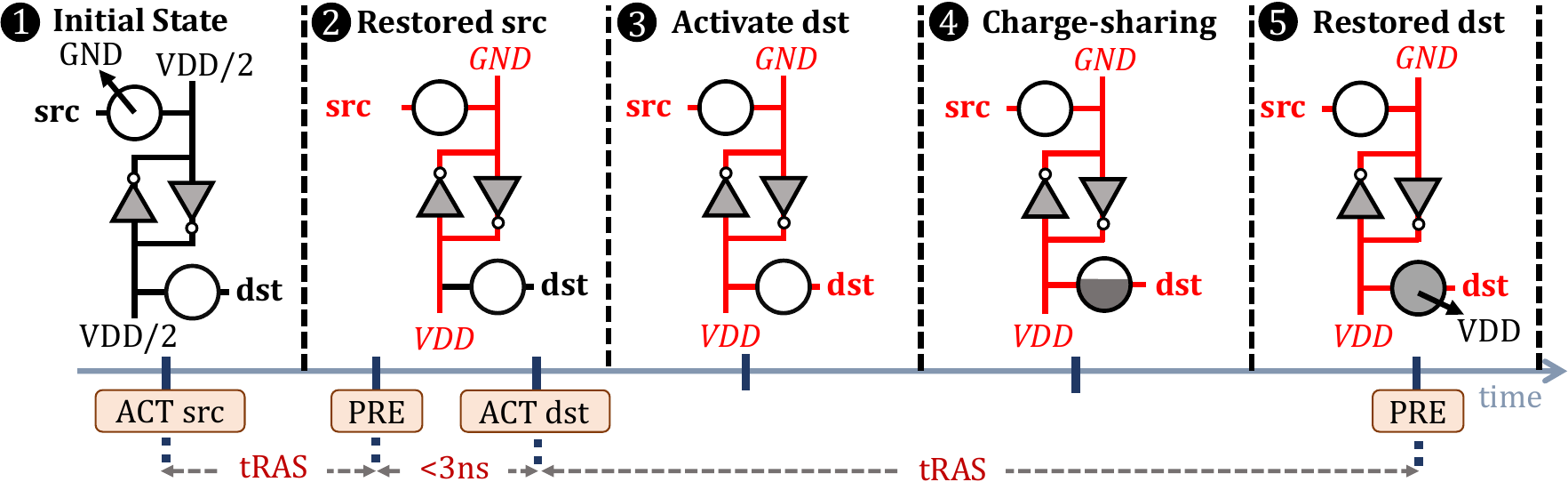}
    \caption{\textbf{Command sequence for performing the NOT operation (dst = NOT(src)) in COTS DRAM chips and the state of cells during each related
    step.} The memory controller issues each command (shown in orange boxes below the time axis) at the corresponding tick mark \omi{on the time axis} and asserted signals are highlighted in red. Cells initially have a voltage level of ground (GND), and the bitline (i.e., \src{}’s bitline) and bitline-bar (i.e., \dst{}’s bitline)
    initially have a voltage level of VDD/2 (\dingOne{}). NOT operation in COTS DRAM chips is performed in four key steps. First, we issue an
    \act{} command to \src{}, i.e., \texttt{ACT}~\src{}, and wait for the manufacturer-recommended \tras{} timing parameter to restore the charge
    of \src{}. As a result, the bitline reaches the \src{} voltage (GND), whereas the bitline-bar reaches the negated \src{} voltage (VDD)
    (\dingTwo{}). Second, we issue a \pre{} command and with violated manufacturer-recommended \trp{} timing, e.g., $<$3ns, we issue another
    \act{} command to activate \dst{}, i.e., \texttt{ACT}~\dst{}. Issuing back-to-back \texttt{PRE} $\rightarrow$ \texttt{ACT}~\dst{} activates
    \dst{} without deactivating \src{} (\dingThree{}) and results in the bitline-bar sharing its charge with \dst{} by driving the negated voltage
    value of \src{} (VDD) into \dst{} (\dingFour{}). Third, we wait for the manufacturer-recommended \tras{} timing parameter, which completely 
    restores the charge of \dst{}, and thus, the negated value of \src{} is written to \dst{} (\dingFive{}). Fourth, we send a \pre{}
    command to complete the process.
    Figure adapted from~\cite{yuksel2024functionallycomplete}.
    }
    \label{fig:fcdram_not}
\end{figure}

\begin{figure}[!htbp]
    \centering
    \includegraphics[width=1\linewidth]{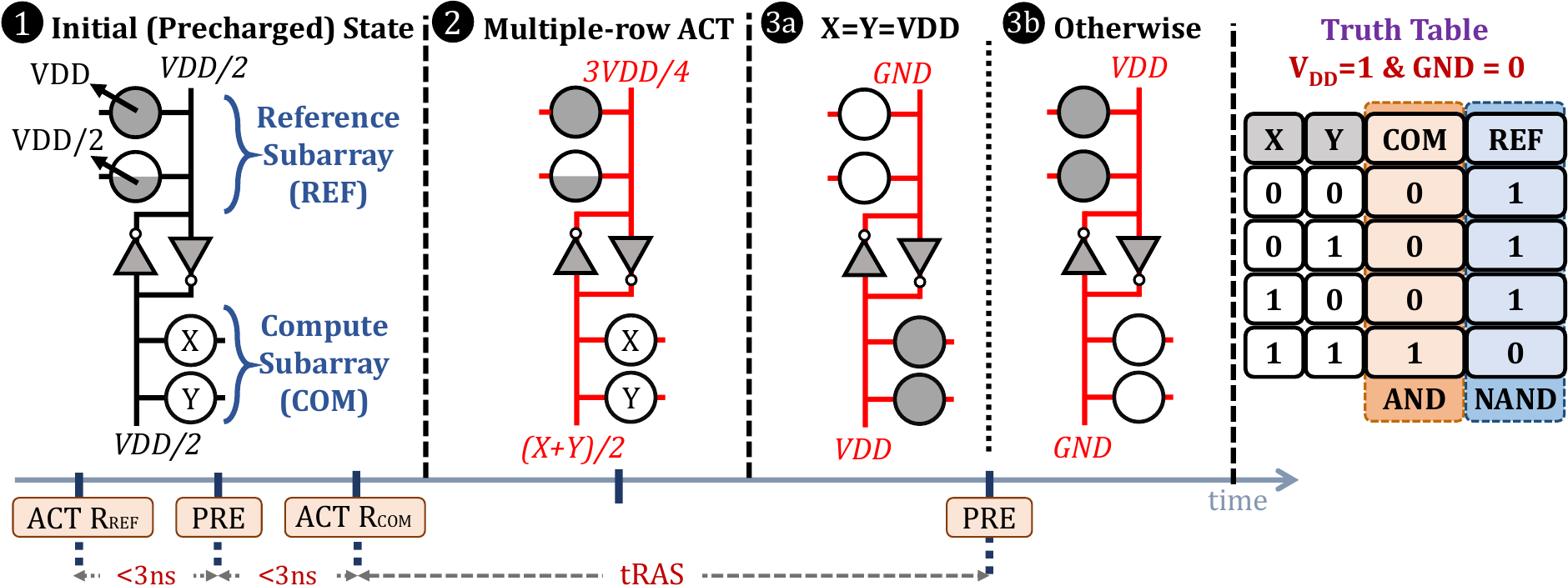}
    \caption{\textbf{Command sequence for performing the two-input AND and NAND operations (i.e., AND(X, Y) and NAND(X, Y)) in COTS DRAM chips and the state of cells during each related step.}
    The memory controller issues each command (shown in orange boxes below the time axis) at the corresponding tick mark and asserted signals are highlighted in red.
    In this figure, we have two neighboring subarrays: the \emph{reference subarray} and the \emph{compute subarray}, each containing two cells. To simplify the explanation, we assume that the bitline has no capacitance (i.e., after charge sharing, the bitline's voltage is the \emph{mean} voltage value stored in DRAM cells that contribute to charge sharing\gf{)}.
    Assume that the \apaAnd{} command sequence with reduced timing simultaneously activates all four rows in these two subarrays where \rref{} points to a row in the reference subarray and \rcom{} points to a row in the compute subarray. Initially, 1) we store VDD in one cell and VDD/2 in the other cell in the reference subarray, and 2) we store a voltage level of X in one cell and Y in the other cell in the compute subarray (\dingOne{}).
    To perform a two-input AND/NAND operation, we first issue one \apaAnd{} command sequence with violated timing parameters (\dingOne{}). Doing so activates four rows simultaneously and enables charge-sharing between their bitlines. At the end of charge-sharing, the reference subarray's bitline voltage (i.e., \vref{})  becomes 3VDD/4 (i.e., the mean of VDD and VDD/2), and the compute subarray's bitline voltage (i.e., \vcom{}) becomes (X+Y)/2 (\dingTwo{}). The sense amplifier then kicks in and amplifies the voltage difference between \vcom{} and \vref{}. If X and Y have VDD (i.e., \vcom{}=VDD), \vcom{} is higher than \vref{}, \omii{which} result\omi{s} in VDD in the compute subarray's activated cells and GND in the reference subarray's activated cells (\circledt{black}{3a}). Otherwise, \vcom{} is lower than \vref{}, \omii{which} result\omi{s} in GND in the compute subarray's activated cells and VDD in the reference subarray's activated cells (\circledt{black}{3b}). After waiting for \tras{}, we issue a \pre{} command to complete the two-input AND/NAND operation. 
    As a result, activated cells in the compute subarray (COM) become the output of the AND(X,Y) operation, and, at the same time, activated cells in the reference subarray (REF) become the output of the NAND(X,Y) operation (as shown in the truth table where X and Y are the inputs and COM and REF are outputs). OR/NOR operations in COTS DRAM chips are similar in nature~\cite{yuksel2024functionallycomplete}.
    Figure adapted from~\cite{yuksel2024functionallycomplete}.
    }
    \label{fig:fcdram_and_example}
\end{figure}

\begin{figure}[!htbp]
    \centering
    \includegraphics[width=1\linewidth]{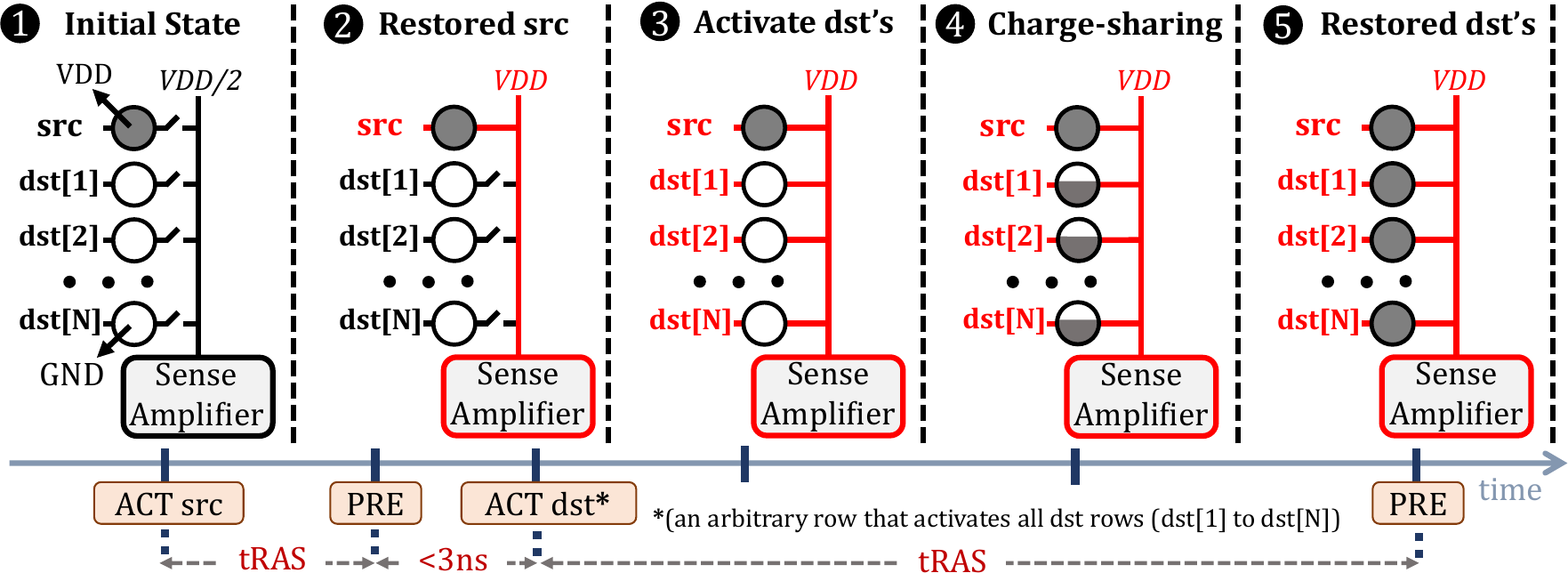}
    \caption{\textbf{Command sequence for performing the Multi-RowCopy operation (i.e., copying src row to N other dst rows simultaneously) in COTS DRAM chips and the state of cells during each related step.} 
    The memory controller issues each command (shown in orange boxes below the time axis) at the corresponding tick mark, and asserted signals are highlighted in red.
    Initially, \src{} cell has VDD, \dst{} cells (i.e., \dst\texttt{[1]} to \dst\texttt{[N]}) have a voltage level of ground (GND) and bitline has a voltage level of VDD/2 (\dingOne{}).
    First, we issue an \act{} command to \src{}, i.e., \texttt{ACT}~\src{}, and wait for the manufacturer-recommended \tras{} timing parameter to restore the charge of \src{}. As a result, the bitline reaches the \src{} voltage (VDD) (\dingTwo{}). Second, we issue a \pre{} command and with violated manufacturer-recommended \trp{} timing, e.g., $<$3ns, we issue another
    \act{} command to activate \dst{}, i.e., \texttt{ACT}~\dst{}. The second \act{} command interrupts the \pre{} command. By doing so, it 1) prevents the bitline from being precharged to VDD/2, 2) keeps \src{} and the sense amplifier enabled, and 3) simultaneously activates \dst{} cells. This results in the bitline sharing its charge with \dst{} cells by driving the voltage value of \src{} (VDD) into \dst{} cells (\dingFour{}). Third, we wait for the manufacturer-recommended \tras{} timing parameter, which results in the sense amplifier overwriting all \dst{} cells with \src{} data (\dingFive{}). Fourth, we send a \pre{}
    command to complete the process.~Multi-RowCopy operation can be used to accelerate not only data copy \& initialization~\cite{seshadri2013rowclone,seshadri2018rowclone} but also cold boot attack prevention as shown in~\cite{yuksel2024simultaneous}.
    }
    \label{fig:simra_multirowcopy}
\end{figure}

\begin{figure}[!htbp]
    \centering
    \includegraphics[width=1\linewidth]{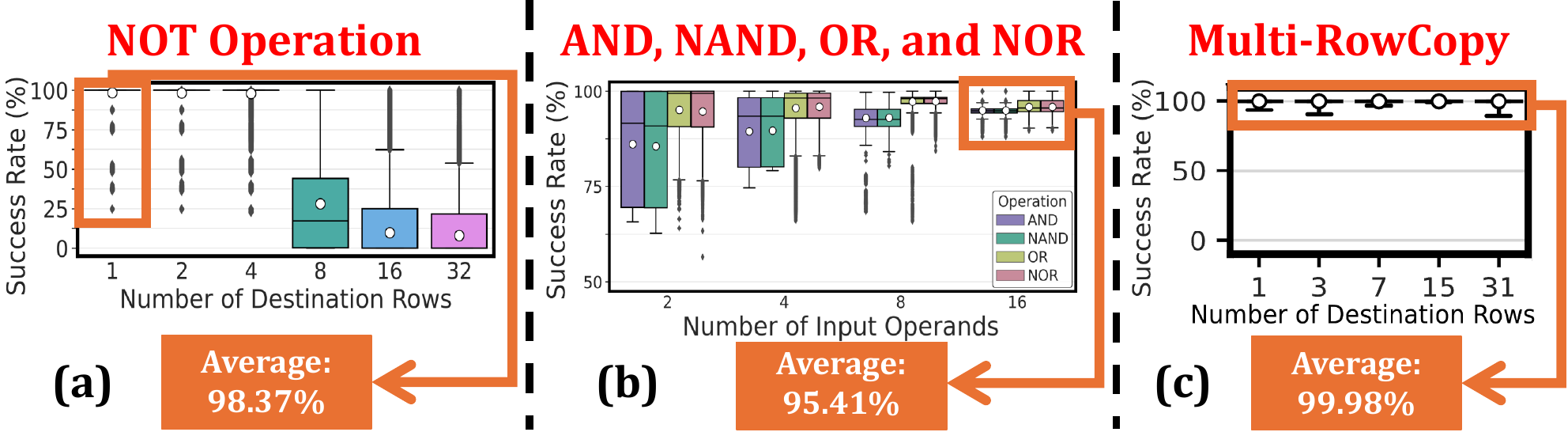}
    \caption{\textbf{Success rates of the NOT operation with varying numbers of destination rows (a) AND, NAND, OR, and NOR operations
    with varying numbers of input operands (b) the Multi-RowCopy operation with varying numbers of destination rows (c), as measured in 224, 224, and 120 COTS DRAM chips, respectively.} On average, we observe a 98.37\% success rate for the NOT operation with one destination row (a), 94.94\%, 94.94\%, 95.85\%, and 95.87\% for 16-input AND, NAND, OR, and NOR operations (b), and 99.98\% for the Multi-Row Copy operation with 31 destination rows (c). We conclude that COTS DRAM chips can execute these operations with high reliability. More results and experimental methodology are in~\cite{yuksel2024functionallycomplete,yuksel2024simultaneous}.}
    \label{fig:fcdram_simra_results}
\end{figure}

A promising line of feasibility study of \gls{pud} systems is to understand the computation capabilities of existing DRAM chips via rigorous experimental testing. Multiple recent works~\cite{yuksel2024functionallycomplete,yuksel2024simultaneous,yuksel2023pulsar} experimentally demonstrate various previously-unknown capabilities in unmodified DRAM chips. These capabilities arise from the operational principles of DRAM circuitry that are exercised by violating the manufacturer-recommended timing parameters~\cite{olgun2023dram, hassan2017softmc}.
In particular, one can simultaneously activate {\em many} DRAM rows in state-of-the-art DRAM chips due to the hierarchical design of the row decoder circuitry~\cite{yuksel2024simultaneous,yuksel2023pulsar,bai2022low,weste2015cmos,turi2008high}.
Exploiting such simultaneous row activation, we~\cite{yuksel2024functionallycomplete,yuksel2024simultaneous,yuksel2023pulsar}
demonstrate that COTS DRAM chips are capable of 1) performing functionally-complete bulk-bitwise Boolean operations: NOT (Fig.~\ref{fig:fcdram_not}), NAND, and NOR, 2) executing up to 16-input AND, NAND (Fig.~\ref{fig:fcdram_and_example}), OR, and NOR operations, and 3) copying the contents of a DRAM row (concurrently) into up to 31 other DRAM rows (Fig.~\ref{fig:simra_multirowcopy}). We evaluate the robustness of these operations across data patterns, temperature, and voltage levels. Our results (Fig.~\ref{fig:fcdram_simra_results}) show that COTS DRAM chips can perform these operations at high success rates ($>$94\%).
These fascinating findings demonstrate the fundamental computation capability of DRAM, even when DRAM chips are {\em not} designed for this purpose, and provide a solid foundation for building new and robust \gls{pud} mechanisms into future DRAM chips and standards.

\begin{figure}[!htbp]
    \centering
    \includegraphics[width=0.9\linewidth]{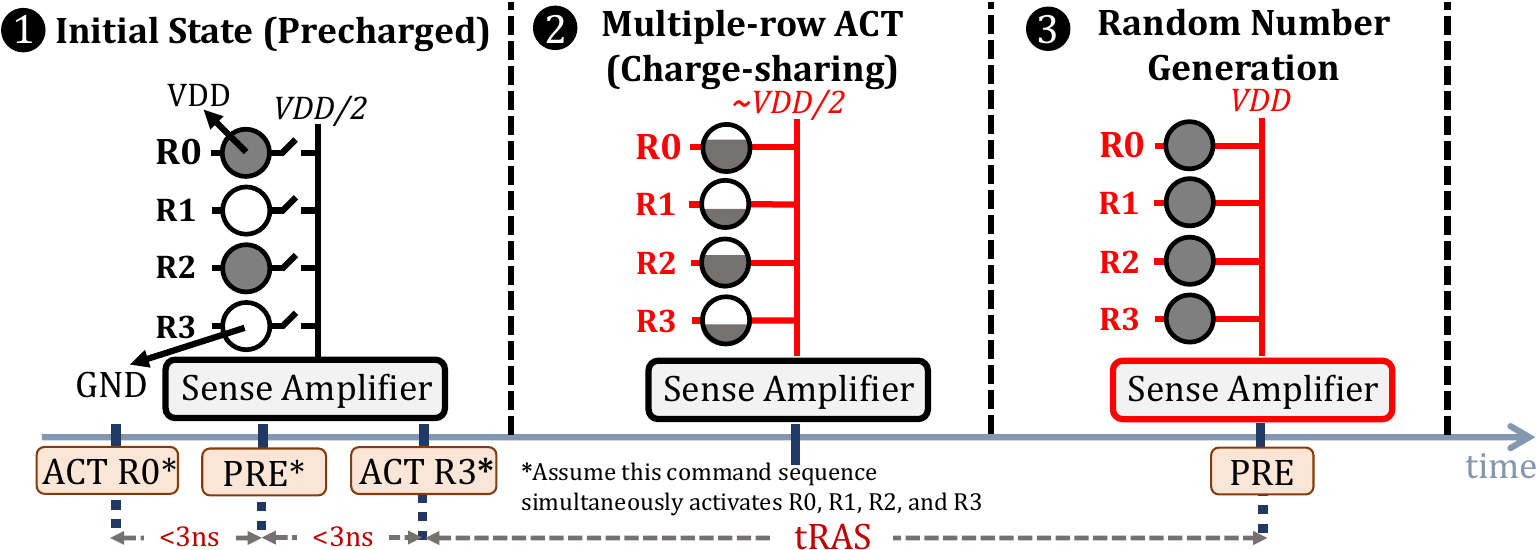}
    \caption{\textbf{Command sequence for true random number generation in COTS DRAM chips and the state of cells during each related step.}
    \iey{The memory controller issues each command (shown in orange boxes below the time axis) at the corresponding tick mark and asserted signals are highlighted in red.}
    Initially, two cells (R0 and R2) have \omii{a voltage level of} VDD, and the remaining two cells (R1 and R3) ground (GND), and bitline has a voltage level of VDD/2 (\dingOne{}).
    To generate random numbers, we first issue one \apatrng{} command sequence (\dingOne{}). Doing so activates four rows simultaneously and enables charge-sharing between their bitlines. As a result, the bitline ends up with a voltage level outside of reliable sensing margins, e.g., $\sim$VDD/2 (\dingTwo{}). The sense amplifier then kicks in and tries to amplify the voltage on the bitline, which results in sampling a random value, e.g., the single depicted bitline is randomly sampled as VDD in this figure (\dingThree{}). Finally, we send a \pre{} command to complete the process. Note that, to be used as TRNG, rows and bitlines need to be profiled~\cite{olgun2021quactrng,kim2019d}.
    }
    \label{fig:mra_trng_mech}
\end{figure}

\begin{figure}[!htbp]
    \centering
    \includegraphics[width=1\linewidth]{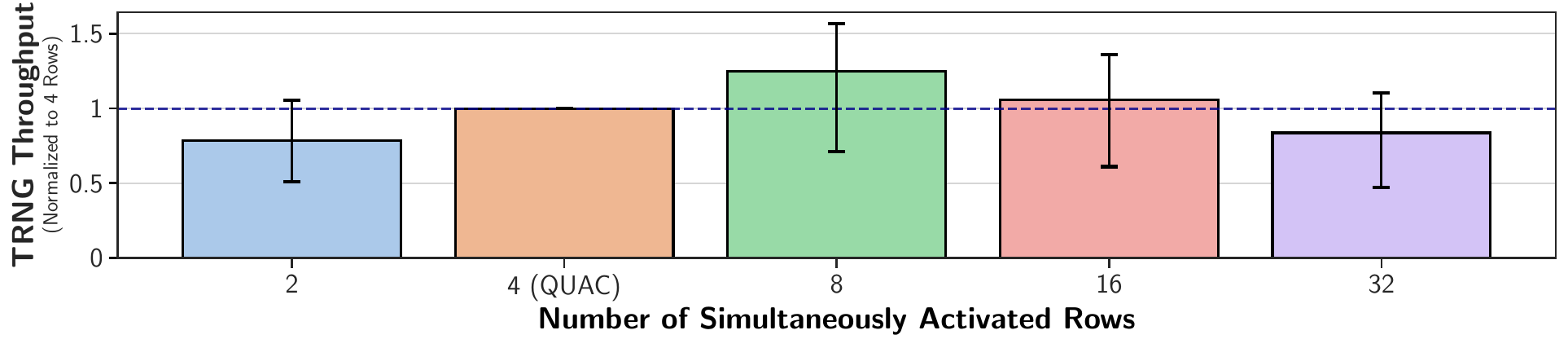}
    \caption{\textbf{Throughput of generating true random numbers, as measured in 96 COTS DRAM chips using multiple-row activation, normalized to state-of-the-art DRAM-based TRNG, QUAC-TRNG (i.e., 4-row activation)~\cite{olgun2021quactrng}.} Each error bar shows the range across all tested chips. We observe that random numbers that are generated with multiple-row activation and then post-processed with the SHA-256 function~\cite{fips2012180} pass \emph{all} NIST STS tests~\cite{rukhin2001statistical}, which means 2-, 4-, 8-, 16-, and 32-row activation generates high-quality true random bitstreams.
    On average, 8- and 16-row activation-based TRNG outperforms the state-of-the-art by 1.25$\times$ and 1.06$\times$, respectively, while 2- and 32-row activation-based TRNG provides 0.69$\times$ and 0.84$\times$ the throughput of the state-of-the-art.
    }  
    \label{fig:mra_trng_results}
\end{figure}

Simultaneous activation of multiple rows in DRAM can be used for generating true random numbers (TRNs) at high throughput (e.g., 3.44 Gb/s per DRAM channel~\cite{olgun2021quactrng}),
widening the workloads supported by \gls{pid} systems (e.g., security-critical workloads) and enabling secure execution support for \gls{pud} systems
that do \emph{not} necessarily have dedicated TRN generation (TRNG) hardware (Fig.~\ref{fig:mra_trng_mech}).
Best prior TRNG using COTS DRAM chips generates TRNs by simultaneously activating four rows~\cite{olgun2021quactrng}. Our ongoing work experimentally studies the simultaneous activation of 2, 8, 16, and 32 rows in a subarray in COTS DRAM chips, showing that 8- and 16-row activation-based TRNG designs provide 1.25$\times$ and 1.06$\times$ higher throughput than the state-of-the-art (Fig.~\ref{fig:mra_trng_results}).
\vspace{8pt}

\vspace{-5pt}
\newpage
\section{Sectored DRAM}
\label{sec:sectored-dram}
\vspace{-2pt}

\begin{figure}[!htbp]
    \centering
    \includegraphics[width=\linewidth]{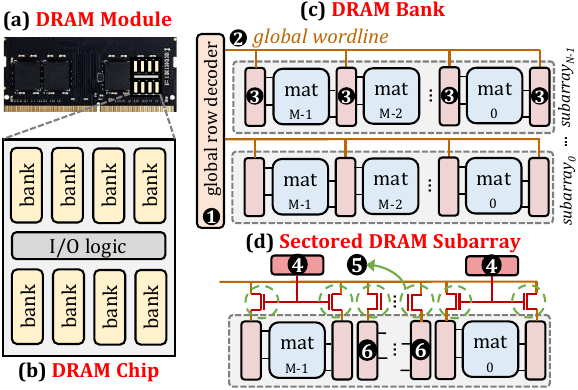}
    \caption{\textbf{A DRAM module (a), a DRAM chip with multiple banks (b), baseline DRAM bank organization with multiple subarrays (c), and a Sectored DRAM subarray (d)~\cite{olgun2024sectored}.} The global row decoder (\dingOne{}) enables a global wordline (\dingTwo{}) based on the higher-order bits of a DRAM row address (not shown). A global wordline enables a local wordline driver (\dingThree{}) that drives a local wordline using the lower-order bits of the DRAM row address (not shown). The baseline DRAM bank activates \omii{\emph{all}} mats \omii{in a subarray} with one activate ($ACT$) DRAM command. Sectored DRAM implements sector latches (\dingFour{}), sector transistors (\dingFive{}), and additional local wordline drivers (\dingSix{}), allowing the DRAM chip to activate any subset of one or more DRAM mats with an $ACT$ command. {To make use of fine-grained DRAM row activation, the memory controller selects sector latches by using the unused bits in the precharge ($PRE$) command's encoding~\cite{jedec2017jedec} to encode the \emph{sector bits}. Each sector bit {encodes if a sector latch is set or reset.} The memory controller sends a bitvector of sector bits with every $PRE$ command. These sector bits are used for the $ACT$ command that follows the $PRE$ command}.}
    \label{fig:secdram_overview}
\end{figure}

Two key coarse-grained access mechanisms lead to wasted energy in modern DRAM chips, also impacting PiM efficiency and programmability: large and fixed-size 
i)~data transfers between DRAM and the memory controller and ii)~DRAM row activations. Sectored DRAM is designed from the ground up as a low-overhead DRAM substrate (Fig.~\ref{fig:secdram_overview}) that reduces wasted energy by enabling fine-grained DRAM data transfer and DRAM row activation. The major idea is to segment the wordlines such that smaller granularity structures are enabled for a given DRAM access. Our results show that Sectored DRAM, compared to a conventional DRAM system, reduces the DRAM energy consumption of data intensive workloads by up to 33\% (20\% on average), while improving performance by up to 36\% (17\% on average). Sectored DRAM's DRAM energy savings combined with its system performance improvement allows system-wide energy savings of up to 23\%.

\glsresetall
\vspace{-2pt}
\section{Conclusion}
\vspace{-2pt}

We highlighted several recent major advances in Processing-in-DRAM, which demonstrate the promising potential of using and enhancing DRAM as a computation substrate. These works also highlight that DRAM (and in general, memory) should be designed, used, and programmed not as an inactive storage substrate, which is {\em business as usual} in modern systems, but instead as a {\em combined computation and storage substrate where both computational capability and storage density are key goals}. Although many challenges remain to enable widespread adoption of Processing-in-DRAM (and Processing-in-Memory in general),
\footnote{Multiple prior works~\cite{ghose.ibmjrd19,mutlu2020modern,mutlu2019processing}
overview these challenges.} 
we believe the mindset and infrastructure shift necessary to enable such a combined computation-storage paradigm remains to be the largest challenge. Overcoming this mindset and infrastructure shift can unleash a fundamentally energy-efficient, high-performance, and sustainable way of designing, using, and programming computing systems.

\section*{Acknowledgments}
{This paper is an extended version of our earlier invited paper and presentation in the ``AI Memory'' focus session of the IEDM 2024 conference~\cite{mutlu2024memory}.} We thank the anonymous reviewers of IEDM 2024{, especially the Memory Technology Committee,} for their encouraging feedback. 
We thank the SAFARI Research Group members for providing a stimulating intellectual {and scientific} environment. We acknowledge the generous gifts from our industrial partners, including Google, Huawei, Intel, and Microsoft. This work{, and our broader work in Processing-in-Memory,} is supported in part by the Semiconductor Research Corporation (SRC), the ETH Future Computing Laboratory (EFCL), and the AI Chip Center for Emerging Smart Systems (ACCESS).

\bibliographystyle{IEEEtran}
\bibliography{references}
\end{document}